\documentclass[journal]{IEEEtran}

\usepackage{amsmath}
\usepackage{subfigure}
\usepackage{cite}
\usepackage{graphicx}
\usepackage{epstopdf}
\usepackage{multirow}
\usepackage{pgfplots}
\pgfplotsset{compat=newest}
\newlength\figureheight
\newlength\figurewidth

\begin{document}

\title{\color{black}{Broadside} \color{black} Dual-channel Orthogonal-Polarization \\ Radiation using a Double-Asymmetric \\ Periodic Leaky-Wave Antenna}

\author{Amar~Al-Bassam,~\IEEEmembership{Student Member,~IEEE,}
        Simon~Otto,~\IEEEmembership{Member,~IEEE,}
				Dirk~Heberling,~\IEEEmembership{Senior~Member,~IEEE,}
        and~Christophe~Caloz,~\IEEEmembership{Fellow,~IEEE}%

\thanks{A.~Al-Bassam and D.~Heberling are with the Institute of High Frequency Technology, RWTH Aachen University, 52062 Aachen, Germany}
\thanks{S.~Otto is with IMST GmbH, 47475 Kamp-lintfort, Germany}%
\thanks{C.~Caloz is with \'Ecole Polytechnique de Montr\'eal, 2500, ch. de
Polytechnique, H3T 1J4, Montr\'eal, Qu\'ebec, Canada.}}
\maketitle%


\begin{abstract}
The paper demonstrates that double unit-cell asymmetry in periodic leaky-wave antennas (P-LWAs), i.e. asymmetry with respect to both the longitudinal and transversal axes of the structure -- or longitudinal asymmetry (LA) \emph{and} transversal asymmetry (TA) -- allows for the simultaneous broadside radiation of two orthogonal modes excited at the two ports of the antenna. This means that the antenna may simultaneously support two orthogonal channels, which represents an interesting polarization diversity characteristics for wireless communications. The double asymmetric (DA) unit cell combines a circularly polarized LA unit cell and a coupled mode TA unit cell, where the former provides equal radiation in the series and shunt modes while the latter separates these two modes in terms of their excitation ports. It is also shown that the degree of TA in the DA unit cell controls the cross-polarization discrimination level. The DA P-LWA concept is illustrated by two examples, a series-fed line-connected patch (SF-LCP) P-LWA and a series-fed capacitively-coupled patch (SF-CCP) P-LWA, via full-wave simulation and also experiment for the SF-LCP P-LWA case.
\end{abstract}


\IEEEpeerreviewmaketitle

\section{Introduction}
\IEEEPARstart{S}{ince} their first explicit report by Hansen~\cite{Hansen1940} in 1946, leaky-wave antennas have been extensively studied and applied due to their attractive properties of frequency scanning and high directivity with feeding simplicity~\cite{Hessel_Collin_book_chap,Tamir_1969,Jackson_Balanis_book_chap,Caloz_McrawHill_2011}.

P-LWAs are advantageously capable of scanning from backfire to endfire, but unfortunately suffer from gain and efficiency instability while scanning through broadside~\cite{Caloz_McrawHill_2011,Jackson_PIEEE_07_2012}. Although this problem, also called the \emph{open-stopband} issue, had been known for decades, it found a solution only very recently. This occurred in the context of transmission-line metamaterials~\cite{Liu_EL_11_2002,Caloz_Wiley_2006}, in terms of the satisfaction of a double condition, the \emph{frequency-balancing} and \emph{$Q$-balancing} condition~\cite{Burghignoli_TAP_2006,Paulotto2009,Otto_TAP_10_2011,Gomez-Diaz2011,Otto_AWPL_06_2012,Otto_TAP_04_2014}.

These findings revealed that the response of P-LWAs is fundamentally dictated by the symmetries of their unit cell, and that mastering symmetry principles resolves the broadside radiation degradation issue and leads to further useful properties. Specifically, a fully symmetric P-LWA always suffers from the broadside radiation degradation issue, while longitudinal asymmetry (LA) or transversal asymmetry (TA) can overcome this issue by ensuring Q-balancing and optimal mode excitation, respectively, in addition to frequency balancing.

Here, we show that combining the aforementioned (LA and TA) asymmetries into a double asymmetric (DA) P-LWA can lead to the simultaneous radiation of two orthogonal modes excited at the two ports of the antenna. Moreover, we show that this principle is universal and illustrate this fact with two different P-LWA structures.

The paper is organized as follows. First, Sec.~\ref{sec:recall} recalls the broadside radiation degradation issue and the concept of P-LWA unit cell asymmetry with its equivalent circuit model. Then Sec.~\ref{sec:DA} combines longitudinal and transversal asymmetries to obtain orthogonal polarizations in opposite propagation directions. Next, two illustrative examples are presented in Sec.~\ref{sec:numerical}, a series-fed line-connected patch (SF-LCP) P-LWA and a series-fed capacitively-coupled patch (SF-CCP) P-LWA. Lastly, Sec.~\ref{sec:matching} designs a specific matching network for the experimental SF-LCP P-LWA, while the complete antenna is demonstrated in Sec.~\ref{sec:exper}.

\section{Fully Symmetric and Single Asymmetric P-LWAs}\label{sec:recall}
In this section, we recall the broadside radiation degradation issue and the recent advances in solving this issue using unit-cell symmetry principles. In the reminder of the paper, we focus on the broadside radiation characteristics of the P-LWA since broadside is the most critical regime of the antenna while off-broadside does not pose any specific problem.

\subsection{Generalities}

\begin{figure}
	\centering
	\includegraphics[width=.48\textwidth]{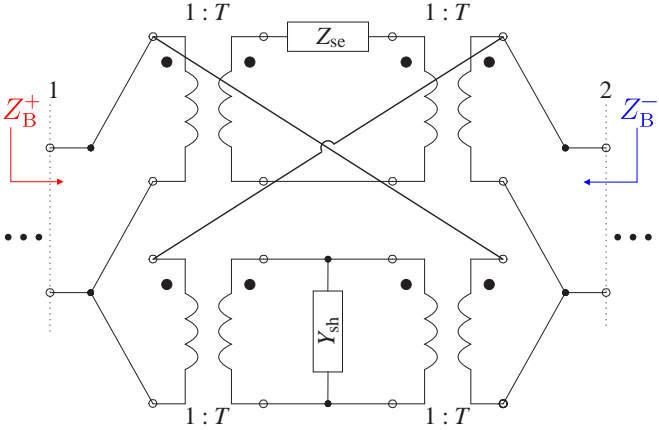}
	\caption{General circuit model for the unit cell of a P-LWA, with impedance $Z_\text{se}$, admittance $Y_\text{sh}$ and four ideal transformers of transformation ratio~$T$~\cite{Otto_TAP_10_2014}.}
	\label{fig:lattice_t}
\end{figure}

\begin{figure}
	\centering
	\includegraphics{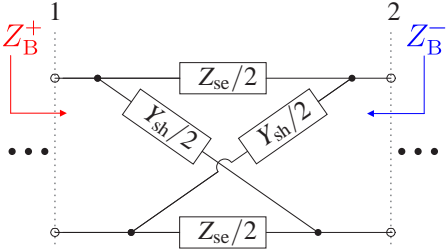}
	\caption{Lattice circuit model for the unit cell of a transversally symmetric P-LWA~\cite{Otto_TAP_10_2011}.}
	\label{fig:lattice}
\end{figure}

The unit cell of a P-LWA may be generally modeled by the equivalent two-port circuit shown Fig.~\ref{fig:lattice_t}, where the series and shunt transmission-line resonators are represented by the impedance $Z_\text{se}$ and the admittance $Y_\text{sh}$, respectively, and where the transformation ratio, $T$, corresponds to the degree of transversal asymmetry~\cite{Otto_TAP_10_2014}. In the case of transversal symmetry, we have $T=1$, and therefore the model of Fig.~\ref{fig:lattice_t} reduces to the equivalent lattice circuit shown in Fig.~\ref{fig:lattice}~\cite{Otto_TAP_10_2011}.

The series and shunt resonators have an approximately linear response around their resonance frequencies. Therefore, under the assumption of frequency balancing, i.e. $\omega_{se}=\omega_\text{sh}=\omega_0$, their immittances around $\omega_0$ may be written~\cite{Otto_TAP_10_2011}

\begin{subequations}
	\begin{align}
		\begin{split}
			Z_\text{se} = R + j2L(\omega-\omega_{0}) = R\left(1+ 2jQ_\text{se}\dfrac{\omega-\omega_0}{\omega_0}\right),
		\end{split}\\
		\begin{split}
			Y_\text{sh} = G + j2C(\omega-\omega_{0}) = G\left(1+ 2jQ_\text{sh}\dfrac{\omega-\omega_0}{\omega_0}\right),
		\end{split}
	\end{align}
	\label{eq:zse_ysh}
\end{subequations}

\noindent where $R$ and $G$ model the loss/radiation contributions of the series and shunt resonators, respectively, and $L$ and $C$ are the series reactance and shunt susceptance, respectively. Furthermore, $Q_\text{se}=\omega_0L/R$ and $Q_\text{sh}=\omega_0C/G$ are the series and shunt quality factors, respectively.

The main results for the DS, the LA and for the TA P-LWAs, that will be next discussed, are summarized in Tab.~\ref{tab:compare}, which also provides off-broadside information for completeness.

\begin{table*}[t]
  \centering
	\caption{Comparison of different symmetry cases of P-LWA's unit cells in terms of Bloch impedance, radiation efficiency and polarization (the unit cell is transversely symmetric for $T=1$ or $\tau=\left(T^2-1/T^2\right)/2=0$)}
	\begin{tabular}{l || c c | c || c c c}
	& \multicolumn{3}{c||}{broadside (0)} & \multicolumn{3}{c}{off-broadside ($\Delta$)}\\
	\cline{2-7}
	& DS & LA & TA & DS & LA & TA \\
	& \includegraphics[width=.1\textwidth]{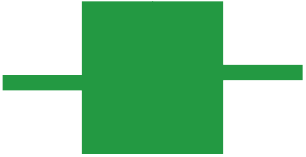} & \includegraphics[width=.1\textwidth]{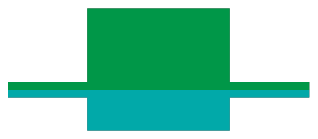} & \includegraphics[width=.1\textwidth]{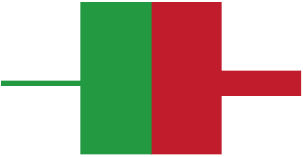} & \includegraphics[width=.1\textwidth]{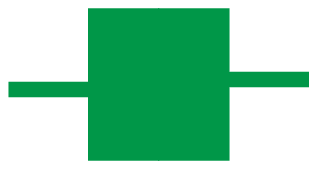} & \includegraphics[width=.1\textwidth]{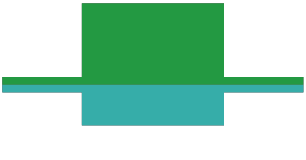} & \includegraphics[width=.1\textwidth]{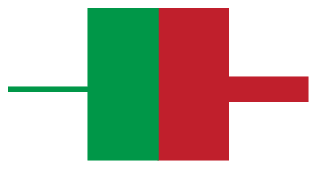} \\

	\hline\hline
	$Z_\text{B}^\pm$  & \multicolumn{2}{c}{$Z_{\text{B},0}^+ =- Z_{\text{B},0}^-$} & \multicolumn{1}{|c||}{$Z_{\text{B},0}^+ \neq Z_{\text{B},0}^-$}  & \multicolumn{3}{c}{$Z_{\text{B},\Delta}^+ = -Z_{\text{B},\Delta}^-$} \\
						 & \multicolumn{2}{c}{$\sqrt{\dfrac{R}{G}}$} & \multicolumn{1}{|c||}{$\mp\dfrac{\tau}{G}+\sqrt{\left(\dfrac{\tau}{G}\right)^2+\dfrac{R}{G}}$} & \multicolumn{3}{c}{$\sqrt{\dfrac{L}{C}}$}\\
						\hline
	$\eta^\pm$ & \multicolumn{2}{c}{$\eta_0^+ = \eta_0^-$} & \multicolumn{1}{|c||}{$\eta_0^+ \neq \eta_0^-$}  & \multicolumn{3}{c}{$\eta_\Delta^+ = \eta_\Delta^-$}\\
	
						 & \multicolumn{2}{c}{$\dfrac{1}{2}\left(\eta_\text{se}+\eta_\text{sh}\right)$} & \multicolumn{1}{|c||}{$\dfrac{\eta_\text{se}RG+\eta_\text{sh}\left|-\tau\pm\sqrt{\tau^2+RG}\right|^2}{RG+\left|-\tau\pm\sqrt{\tau^2+RG}\right|^2}$} & \multicolumn{3}{c}{$\dfrac{\eta_\text{se}Q_\text{sh}+\eta_\text{sh}Q_\text{se}}{Q_\text{sh}+Q_\text{se}}$} \\
						\hline

	Polarization & linear & elliptical & linear & linear & elliptical & linear
  \end{tabular}
	\label{tab:compare}
\end{table*}

\subsection{Double Symmetric (DS) Case}\label{sec:DS}

It has been shown in~\cite{Otto_TAP_10_2011} that a P-LWA that is symmetric with respect to its two axes, i.e. double symmetric (DS), suffers from the broadside radiation degradation issue, with a radiation efficiency theoretically limited to 50\% at broadside~\cite{Otto_AWPL_06_2012}. This issue is caused by equal power splitting between the series and shunt resonators, $P_\text{se}=|I_\text{se}|^2R=P_\text{sh}=|V_\text{sh}|^2G$, where the contribution to broadside radiation from the shunt resonator is null due antisymmetric field cancellation and radiation contribution is therefore only from the series resonator. Moreover, since only the series resonator radiates, the overall polarization of the antenna is exclusively linear in the broadside direction.

\subsection{Longitudinal Asymmetric (LA) Case}\label{sec:LA}

Introducing longitudinal asymmetry (LA) to the unit cell of a P-LWA affects only its shunt resonator. Increasing LA decreases the $Q$-factor of this resonator and therefore increases its radiation contribution. Since the unit cell is LA, the DS radiation cancellation at broadside is mitigated and therefore the overall broadside radiation efficiency increases~\cite{Otto_AWPL_06_2012, Otto_GeMIC_03_2012}. Furthermore, it was found in~\cite{Otto_TAP_04_2014} that the series and shunt modes are in a quadrature phase relationship, so that the polarization is generally elliptical, and circular if the radiation contributions of the two modes are equal in magnitude.

\subsection{Transversal Asymmetric (TA) Case}\label{sec:TA}

It was pointed out in Sec~\ref{sec:DS} that DS in P-LWA implies radiation efficiency limitation to $50~\%$ due to the non-radiation of the shunt resonator. It was next mentioned, in Sec~\ref{sec:LA}, that the efficiency of the shunt resonator can be increased, thereby increasing the overall efficiency, but at the cost of loosing polarization linearity. The TA P-LWA solves the broadside issue without compromising polarization, hence featuring high overall efficiency and linear polarization~\cite{Otto_TAP_10_2014}, by exciting the series and shunt resonators in optimal power ratios, which is allowed by the fact that the TA structure breaks the equal power splitting condition.

\section{Double Asymmetric (DA) P-LWA}\label{sec:DA}

\begin{figure}
	\centering
	\includegraphics[]{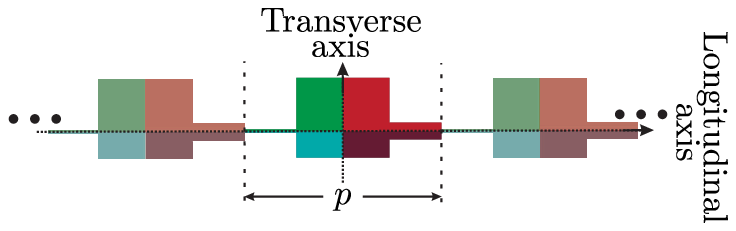}
	\caption{\color{black} Generic representation of a DA (double asymmetric) P-LWA. The unit cell is asymmetric w.r.t. both the transversal and longitudinal (propagation) axes.}
	\label{fig:DA_def}
	\end{figure}

This section introduces the asymmetry configuration combining the asymmetries in Sec.~\ref{sec:LA} (LA) and Sec.~\ref{sec:TA} (TA)\color{black}, as shown in Fig.~\ref{fig:DA_def}\color{black}. As will be shown, this double asymmetric (DA) configuration combines properties inherited from both LA and TA, where LA provides high efficiency in both the series and shunt resonators while TA controls the series and shunt power ratios at broadside.

The TA case has been investigated in~\cite{Otto_TAP_10_2014}, using the transformer-based equivalent circuit shown in Fig.~\ref{fig:lattice_t}. It was shown that the degree of TA can be modeled solely by the transformation ratio $T$, which controls the level of coupling between the series and shunt resonators. Furthermore, it was shown that the shunt-to-series power ratios at broadside \color{black}($\omega-\omega_0 = 0$) \color{black} as seen from opposite directions, i.e. forward~($+$) and backward~($-$) directions, are proportional to the corresponding $T$-dependent Bloch impedances\footnote{\color{black}For the sake of self-consistency, Appendix~\ref{appen:network} concisely derives and analyses the formulas for the Bloch impedance.\color{black}} at broadside,

\begin{equation}\label{eq:ZB}
Z_{\text{B},0}^\pm(T)=\mp\dfrac{\tau}{G}+\sqrt{\left(\dfrac{\tau}{G}\right)^2+\dfrac{R}{G}},
\quad
\tau=\dfrac{1}{2}\left(T^2-1/T^2\right),
\end{equation}

\noindent as

\begin{align}
		\left(\dfrac{P_\text{sh}}{P_\text{se}}\right)^\pm=\dfrac{G}{R}|Z_{\text{B},0}^\pm|^2(T),
		\label{eq:P_ratio}
\end{align}

\noindent where the transformation ratio is obtained from the $ABCD$-parameters as $T = \sqrt{(D+1)/(A+1)}$. If $T$ is unity, or equivalently $\tau=0$, then Eq.~\eqref{eq:ZB} yields $Z_{\text{B},0}^\pm(T=1)=\pm\sqrt{R/G}$, and therefore the power equally splits between the series and shunt resonators, which corresponds to the DS or LA cases (Tab.~\ref{tab:compare}). Furthermore, Eq.~\eqref{eq:P_ratio} reveals that the forward and backward Bloch impedances control the corresponding power ratios, which will ultimately determine the corresponding radiation efficiencies given as~\cite{Otto_TAP_10_2014}

\begin{align}
	\eta^\pm = \dfrac{\eta_\text{se}R+\eta_\text{sh}G\left|Z_\text{B}^\pm\right|^2}{R+G\left|Z_\text{B}^\pm\right|^2},
	\label{eq:eta_pm}
\end{align}

\noindent where $\eta_\text{se}$ and $\eta_\text{sh}$ are the series and shunt radiation efficiencies, respectively.

In the DA case, both the series and shunt resonators have high efficiency, due to LA, since the LA properties were derived in complete generality, irrespectively of whether the unit cell is transversally symmetric or asymmetric (Sec.~\ref{sec:LA}). In fact, TA does not affect the intrinsic series and shunt resonators, but only their coupling, as seen by comparing Figs.~\ref{fig:lattice_t} and~\ref{fig:lattice}.

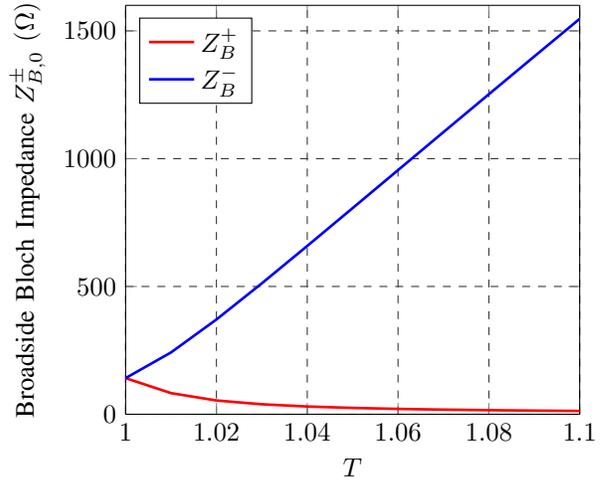
\begin{figure}
		\centering
		\hspace{5mm}
		\setlength\figureheight{.3\textwidth}
		\setlength\figurewidth{.35\textwidth}
%
%
\begin{tikzpicture}[%
trim axis left,trim axis right,/pgf/number format/.cd, 1000 sep={}
]
\pgfplotsset{grid style={dashed,black!80!white}}

\begin{axis}[%
width=0.951\figurewidth,
height=\figureheight,
at={(0\figurewidth,0\figureheight)},
scale only axis,
separate axis lines,
every outer x axis line/.append style={black},
every x tick label/.append style={font=\color{black}},
xmin=1,
xmax=1.1,
xlabel={$T$},
xmajorgrids,
every outer y axis line/.append style={black},
every y tick label/.append style={font=\color{black}},
ymin=0,
ymax=1600,
ylabel={Broadside Bloch Impedance $Z_{B,0}^\pm~(\Omega)$},
ymajorgrids,
axis background/.style={fill=white},
legend style={at={(0.03,0.97)},anchor=north west,legend cell align=left,align=left,draw=black},
legend style={fill=white}
]
\addplot [color=red,solid,line width=1.0pt]
  table[row sep=crcr]{%
1	141.42135623731\\
1.01	82.6801696362689\\
1.02	53.9295211421521\\
1.03	39.0427418334486\\
1.04	30.3699765680807\\
1.05	24.7931257879017\\
1.06	20.9347579206378\\
1.07	18.1165712167829\\
1.08	15.9716803057338\\
1.09	14.2860603423004\\
1.1	12.9270621565613\\
};
\addlegendentry{$Z_B^+$};

\addplot [color=blue,solid,line width=1.0pt]
  table[row sep=crcr]{%
1	141.42135623731\\
1.01	241.895972008586\\
1.02	370.85439619021\\
1.03	512.259105298431\\
1.04	658.545124497076\\
1.05	806.675211955702\\
1.06	955.348997863679\\
1.07	1103.96165812394\\
1.08	1252.2163990986\\
1.09	1399.96608727606\\
1.1	1547.14193818962\\
};
\addlegendentry{$Z_B^-$};

\end{axis}
\end{tikzpicture}%
		\caption{Bloch impedance at broadside versus $T$ (degree of TA) for $R=10~\Omega$ and $G=500~\mu S$ [Eq.~\eqref{eq:ZB}].}
		\label{fig:ZB_vs_T}
	\end{figure}

At the same time, the Bloch impedance strongly depends on the direction of propagation for high TA ($T\gg1$), as illustrated in Fig.~\ref{fig:ZB_vs_T}, with

\begin{subequations}
	\begin{align}
		\begin{split}
			\left.Z_{B}^+\right|_{T\gg1}\rightarrow 0,
\label{eq:ZB_extrema}
		\end{split}\\
		\begin{split}
			\left.Z_{B}^-\right|_{T\gg1}\rightarrow \infty.
\label{eq:ZB_extremb}
		\end{split}
	\end{align}
	\label{eq:ZB_extrem}
\end{subequations}

\noindent As a result, from~\eqref{eq:eta_pm}, we have

\begin{subequations}
	\begin{align}
		\begin{split}
			\left.\eta^+\right|_{T\gg1} \approx \eta_\text{se},
	\label{eq:eta_extrema}
		\end{split}\\
		\begin{split}
			\left.\eta^-\right|_{T\gg1} \approx \eta_\text{sh},
	\label{eq:eta_extremb}
		\end{split}
	\end{align}
	\label{eq:eta_extrem}
\end{subequations}

\noindent which shows that exciting a DA P-LWA with high TA corresponds to selectively exciting the series resonator in the forward direction and the shunt resonator in the backward direction. Moreover, from LA, both resonators can reach maximal efficiency (100$\%$ in the absence of dissipation loss), which means that maximal efficiency can be achieved from both ports. Finally, since the series and shunt resonators feature mutually orthogonal fields~\cite{Otto_TAP_04_2014}, the polarizations from the two ports will be orthogonal to each other.

Specifically, the series polarization (forward direction) will be longitudinal while the shunt polarization (backward direction) will be transverse.

\section{Full-Wave Analysis}\label{sec:numerical}

To demonstrate and provide further insight into the proposed dual-channel orthogonal-polarization DA P-LWA, this section provides the full-wave analysis of two DA P-LWAs with different geometrical parameters, a series-fed connected-line patch (SF-LCP) P-LWA and a series-fed capacitively-coupled patch (SF-CCP) P-LWA, designed for a transition frequency of 24~GHz.

\subsection{Series-Fed Line-Connected Patch (SF-LCP) P-LWA}\label{sec:SFP}

\begin{figure}
	\centering
	\includegraphics[width=.48\textwidth]{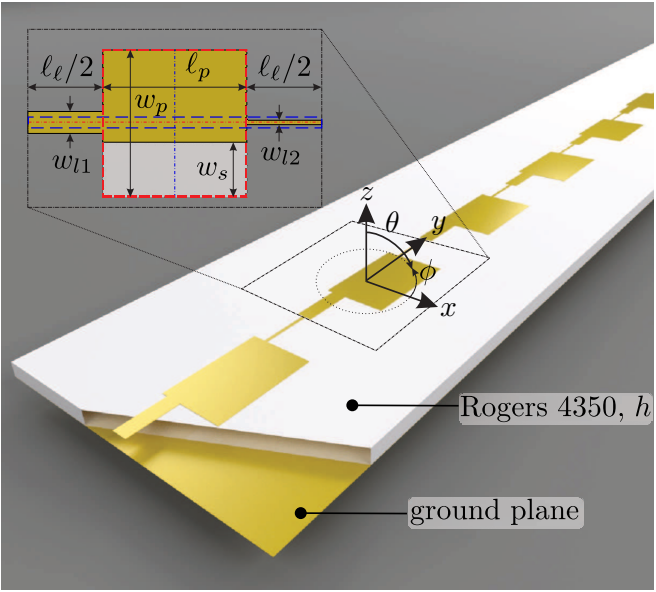}
	\caption{Series-fed line-connected patch (SF-LCP) DA P-LWA. The corresponding DS structure is shown by dashed lines in the inset. Symmetry w.r.t. the longitudinal axis ($y$-axis) is broken by trimming the patch by a quantity $w_s$, while symmetry w.r.t. the transverse axis ($x$-axis) is broken by changing the width of the feeding line sections to the two  widths $w_{l1}$ and $w_{l2}$.}
	\label{fig:SFP_layout}
\end{figure}

The SF-LCP P-LWA, which radiates in the $n=-1$ space harmonic, is shown in Fig.~\ref{fig:SFP_layout}. The structure consists in a line-patch patterned metallization layer on a grounded Rogers RO~4350 substrate with height $h=508~\mu\textrm{m}$. The dimensions of the unit-cell are given in Table~\ref{tab:SFP_dim}.

The design is performed in three steps. First, starting from the DS configuration, the unit cell is optimized to satisfy the frequency-balancing condition, $\omega_\text{se}=\omega_\text{sh}=\omega_{0}$, by tuning the length parameters $l_l$ (feeding line length) and $l_p$ (patch length). \color{black} The frequency-balancing optimization is covered in~\cite{Caloz_Wiley_2006,Jackson_Balanis_book_chap,Otto_TAP_10_2011}\color{black}. Then, $Q$-balancing is achieved by tuning the width $w_s$ (patch width from one side), which leads to LA, and therefore elliptical polarization. \color{black} $Q$-balancing optimization has been extensively described in~\cite{Otto_TAP_04_2014}. Note that parameter $w_s$ does not influence the transversal asymmetry and therefore does not affect $T$\color{black}. Finally, TA, and thereby DA, is added by making the widths of the feeding lines, $w_{l1}$ and $w_{l2}$, different from each other, and tuning them properly. The frequency balancing condition is typically altered through these steps, which are therefore iteratively repeated until the design is fully satisfactory. \color{black} The design is considered complete when the cross-polarization discrimination level, controlled by tuning $T$, is sufficiently high and when the  broadside directivities at the two ports are sufficiently close, which is ensured by $Q$-balancing.\color{black}

\begin{table}
	\centering
		\caption{Dimensions in mm of the DA SF-LCP P-LWA shown in Fig.~\ref{fig:SFP_layout}.}
		\begin{tabular}{c|c|c|c|c|c}
			$l_{l}$ & $l_{p}$ & $w_{l1}$ & $w_{l2}$ & $w_{p}$ & $w_{s}$ \\ \hline \hline
			3.41 & 3.28 & 0.10 & 0.50 & 3.30 & 1.20
		\end{tabular}
	\label{tab:SFP_dim}
\end{table}

\begin{figure}
		\centering
		\hspace{5mm}
		\setlength\figureheight{.25\textwidth}
		\setlength\figurewidth{.35\textwidth}
		\input{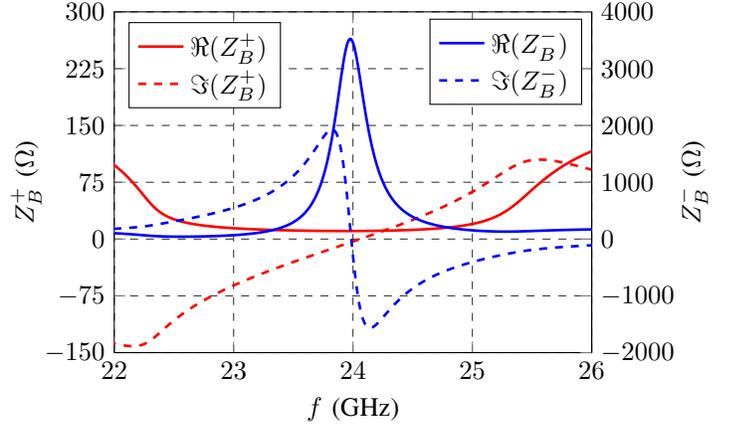}
		\caption{Bloch impedance of the forward and backward propagating waves, $Z_\text{B}^+$ and $Z_\text{B}^-$, respectively, for $T\approx 1.3$. $Z_\text{B}^+$ exhibits the behaviour of a series resonator (low impedance) and $Z_\text{B}^-$ of a shunt resonator (high impedance).}
		\label{fig:ZB_SFP}
	\end{figure}

The Bloch impedances for the forward direction ($Z_\text{B}^+$) and backward direction ($Z_\text{B}^-$) are plotted in Fig.~\ref{fig:ZB_SFP}. $Z_\text{B}^+$ exhibits a low impedance at broadside, which indicates the behavior of a series resonating circuit, while $Z_\text{B}^-$ exhibits a high impedance corresponding to a shunt resonating circuit, as understood from~\eqref{eq:zse_ysh} with $R\rightarrow 0$ and $G\rightarrow 0$, respectively. As indicated in Sec.~\ref{sec:DA} and verified in Eq.~\eqref{eq:P_ratio}, different Bloch impedances lead to unequal series and shunt power split ($P_\text{se}\neq P_\text{sh}$) at broadside. Specifically, in the high TA regime ($T\gg 1$), $Z_\text{B}^+\rightarrow 0$ [Eq.~\eqref{eq:ZB_extrema}], and therefore $P_\text{sh}\rightarrow 0$ [Eq.~\eqref{eq:eta_extrema}], so that only the series resonator is excited; conversely, $Z_\text{B}^-\rightarrow \infty$ [Eq.~\eqref{eq:ZB_extremb}], and therefore $P_\text{se}\rightarrow 0$ [Eq.~\eqref{eq:eta_extrema}], so that only the shunt resonator is excited. So, we have longitudinal polarization in the forward direction and transverse polarization in the backward direction.

\begin{figure}
	\centering
	\setlength\figureheight{5cm}
	\setlength\figurewidth{5cm}
	\subfigure[]{\includegraphics[]{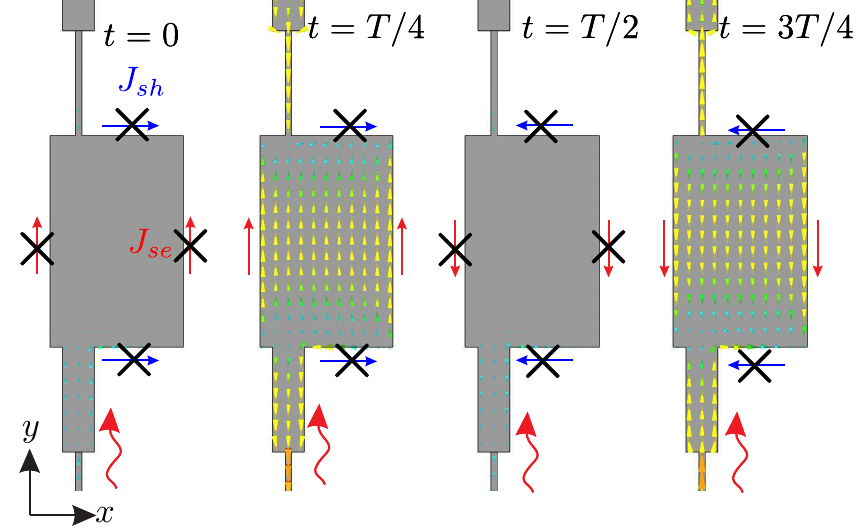}\label{fig:Jsurf_SFP_se.eps}}
	\subfigure[]{\includegraphics[]{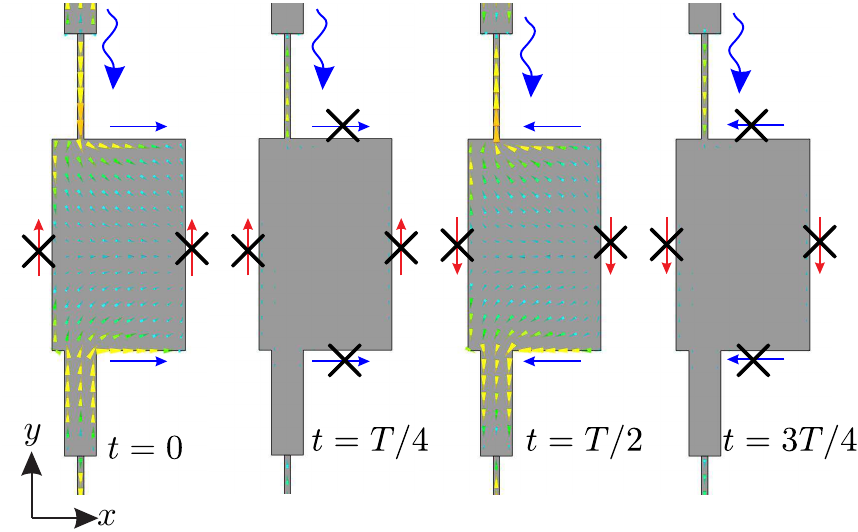}\label{fig:Jsurf_SFP_sh.eps}}
	\caption{Surface current density evolution obtained by full-wave simulation on the top metallization of the DA SF-LCP LWA unit cell. (a) Excitation in the forward direction corresponds to field distribution of the series mode for $t=T/4$ and $t=3T/4$ and (b) excitation from the backward direction corresponds to field distribution of the shunt mode for $t=0$ and $t=T/2$.}
	\label{fig:Jsurf_SFP}
\end{figure}

Figure~\ref{fig:Jsurf_SFP} plots the surface current density on the top metallization of the SF-LCP P-LWA at the time instants $t=0$, $t=T/4$, $t=T/2$ and $t=3T/4$, where $T$ represents here the period of the wave (and not the degree of transversal asymmetry). Figure~\ref{fig:Jsurf_SFP_se} corresponds to the case of forward excitation. The surface currents are approximately zero $t=0$ and $t=T/2$, while they are nonzero, longitudinal and maximal at the long edges of the patch at $t=T/4$ and $t=3T/4$, which naturally corresponds, as expected, to longitudinal polarization. The backward case is presented in Fig.~\ref{fig:Jsurf_SFP_sh}. This time, the surface currents are approximately zero at $t=T/4$ and $t=3T/4$, while they are nonzero, transverse and maximal at the short edges of the patch at  $t=0$ and $t=T/2$, which naturally corresponds, as expected, to longitudinal polarization. Interestingly, the series and shunt resonances are in quadrature phase relation, as it can be seen by comparing Fig~\ref{fig:Jsurf_SFP_se} to Fig~\ref{fig:Jsurf_SFP_sh}.

Figure~\ref{fig:D_vs_T} shows the directivity dependency on the degree of transversal asymmetry ($T$), for the SF-LCP P-LWA being excited in the forward ($+\hat{y}$) direction, i.e. from Port~1, and in the backward  ($-\hat{y}$) direction, i.e. from Port~2. The curves at the top correspond to co-polarization \color{black} (transverse/vertical $v$ for Port~1 and longitudinal/horizontal $h$ for Port~2) \color{black}and those at the bottom correspond to cross-polarization \color{black} ($h$ for Port~1 and $v$ for Port~2)\color{black}. These results were obtained by FDTD full-wave simulation by varying the widths of the transmission lines, while maintaining the double balancing condition, under the constraint $w_{l1}\neq w_{l2}$.

\begin{figure}
	\centering
	\hspace{5mm}
	\setlength\figureheight{.3\textwidth}
	\setlength\figurewidth{.35\textwidth}
%
%
\begin{tikzpicture}[%
trim axis left,trim axis right,/pgf/number format/precision=2,/pgf/number format/fixed
]
\pgfplotsset{grid style={dashed,black}}

\begin{axis}[%
width=0.951\figurewidth,
height=\figureheight,
at={(0\figurewidth,0\figureheight)},
scale only axis,
xmin=1,
xmax=1.3,
xtick={   1, 1.05,  1.1, 1.15,  1.2, 1.25,  1.3},
xlabel={$T$},
xmajorgrids,
ymin=-10,
ymax=20,
ytick={-10,-5,0,5,10,15,20},
ylabel={Directivity $D(\phi=90^\circ,\theta=0^\circ)$ (dBi)},
ymajorgrids,
axis background/.style={fill=white},
legend style={at={(0.97,0.5)},anchor=east,legend cell align=left,align=left,draw=white!15!black},
legend style={fill=white}
]
\addplot [color=red,solid,line width=1.3pt]
  table[row sep=crcr]{%
1.00003886164028	14.00726119468\\
1.01257577144587	14.7526423047203\\
1.0229987394275	15.1778154725574\\
1.03669290599983	15.376981203685\\
1.04952662306364	15.3143175056457\\
1.06157877160113	15.1257924045583\\
1.07635051623125	14.7942927309095\\
1.0893956017647	14.4726134543923\\
1.10295576893376	14.1339693999738\\
1.11604138041654	13.7984053139878\\
1.12927345508248	13.4710177440481\\
1.14292447935682	13.1576930090792\\
1.15617514065785	12.8565638624099\\
1.17529087784906	12.4492583215497\\
1.18974703772829	12.1644967766022\\
1.2052370259554	11.8921561117147\\
1.221338185652	11.6293928111672\\
1.23808341607765	11.3783017268405\\
1.25865764038518	11.0702947816643\\
1.27656684526565	10.8298328149195\\
1.29539899983609	10.5975085044381\\
};
\addlegendentry{Port 1: \textcolor{black}{$D_{v,0}$}};

\addplot [color=blue,dashed,line width=1.3pt]
  table[row sep=crcr]{%
1.00003886164028	14.0803627677385\\
1.01257577144587	12.9300115273348\\
1.0229987394275	11.6559180081913\\
1.03669290599983	9.87703275070231\\
1.04952662306364	8.16342157767258\\
1.06157877160113	6.5396925718988\\
1.07635051623125	4.68049140058559\\
1.0893956017647	3.20605265787952\\
1.10295576893376	1.78174823787504\\
1.11604138041654	0.555902867692225\\
1.12927345508248	-0.558251479910894\\
1.14292447935682	-1.57187484552099\\
1.15617514065785	-2.49281270488217\\
1.17529087784906	-3.56189210436287\\
1.18974703772829	-4.31722623979846\\
1.2052370259554	-4.99894726314329\\
1.221338185652	-5.67320936689293\\
1.23808341607765	-6.22672689779364\\
1.25865764038518	-6.72151719343861\\
1.27656684526565	-7.15035839439633\\
1.29539899983609	-7.53464624549775\\
};
\addlegendentry{Port 1: \textcolor{black}{$D_{h,0}$}};

\addplot [color=red,dashed,line width=1.3pt]
  table[row sep=crcr]{%
1.00003886164028	14.0077463238691\\
1.01257577144587	12.7351431383838\\
1.0229987394275	11.5103906502715\\
1.03669290599983	9.75363106322698\\
1.04952662306364	8.04970289388172\\
1.06157877160113	6.43476987130742\\
1.07635051623125	4.49095343588999\\
1.0893956017647	3.02618910333251\\
1.10295576893376	1.66687962401596\\
1.11604138041654	0.484278004624173\\
1.12927345508248	-0.572560726904726\\
1.14292447935682	-1.5276527133297\\
1.15617514065785	-2.3844424926981\\
1.17529087784906	-3.46520682021452\\
1.18974703772829	-4.18458527991453\\
1.2052370259554	-4.83716840493855\\
1.221338185652	-5.43746221413386\\
1.23808341607765	-5.98565044763006\\
1.25865764038518	-6.67448013203961\\
1.27656684526565	-7.15474447974304\\
1.29539899983609	-7.60176942378291\\
};
\addlegendentry{Port 2: \textcolor{black}{$D_{v,0}$}};

\addplot [color=blue,solid,line width=1.3pt]
  table[row sep=crcr]{%
1.00003886164028	14.0798635384991\\
1.01257577144587	14.9706872723865\\
1.0229987394275	15.3904145730682\\
1.03669290599983	15.5989307500968\\
1.04952662306364	15.5594820629631\\
1.06157877160113	15.3871381599669\\
1.07635051623125	15.081456569602\\
1.0893956017647	14.7698845128235\\
1.10295576893376	14.4326076938896\\
1.11604138041654	14.1057746851278\\
1.12927345508248	13.7869721953745\\
1.14292447935682	13.4827901798871\\
1.15617514065785	13.1910944081903\\
1.17529087784906	12.8011478994121\\
1.18974703772829	12.5257259415547\\
1.2052370259554	12.26315799023\\
1.221338185652	12.0096489944379\\
1.23808341607765	11.7717410312342\\
1.25865764038518	11.4823444271025\\
1.27656684526565	11.2547626471267\\
1.29539899983609	11.0347073688234\\
};
\addlegendentry{Port 2: \textcolor{black}{$D_{h,0}$}};

\end{axis}
\end{tikzpicture}%
	\caption{Broadside directivity \color{black}(at $\theta = 0^\circ$ and $\phi = 90^\circ$) \color{black} versus of degree of transversal asymmetry ($T$) obtained by full-wave simulation for $8$ unit cells. In the case $T=1$, the unit cell is transversally symmetric and therefore elliptically polarized.}
	\label{fig:D_vs_T}
\end{figure}
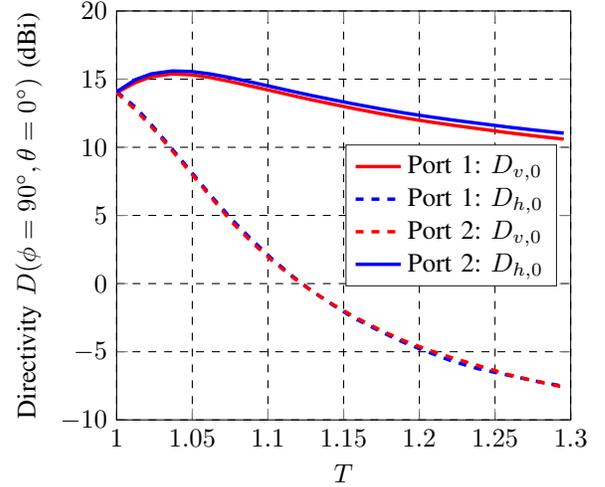

\begin{figure}
	\centering
	\hspace{5mm}
	\setlength\figureheight{.2\textwidth}
	\setlength\figurewidth{.35\textwidth}
%
%
\begin{tikzpicture}[%
trim axis left,trim axis right,/pgf/number format/precision=2,/pgf/number format/fixed
]
\pgfplotsset{grid style={dashed,black}}

\begin{axis}[%
width=0.951\figurewidth,
height=\figureheight,
at={(0\figurewidth,0\figureheight)},
scale only axis,
xmin=1,
xmax=1.3,
xtick={1, 1.05,  1.1, 1.15,  1.2, 1.25,  1.3},
xlabel={$T$},
xmajorgrids,
ymin=0.6,
ymax=0.9,
ylabel={Radiation Efficiency $\eta$},
ymajorgrids,
axis background/.style={fill=white},
legend style={at={(0.97,0.1)},anchor=south east,legend cell align=left,align=left,draw=white!15!black},
legend style={fill=white}
]
\addplot [color=blue,solid,line width=1.3pt]
  table[row sep=crcr]{%
1.00003886164028	0.786912995\\
1.01257577144587	0.765898651\\
1.0229987394275	0.766712339\\
1.03669290599983	0.76473175\\
1.04952662306364	0.762393361\\
1.06157877160113	0.762168795\\
1.07635051623125	0.763711901\\
1.0893956017647	0.764357784\\
1.10295576893376	0.767352909\\
1.11604138041654	0.769967545\\
1.12927345508248	0.772751875\\
1.14292447935682	0.77559433\\
1.15617514065785	0.778892379\\
1.17529087784906	0.785969953\\
1.18974703772829	0.789852052\\
1.2052370259554	0.793681139\\
1.221338185652	0.797728865\\
1.23808341607765	0.801999427\\
1.25865764038518	0.808159657\\
1.27656684526565	0.812654881\\
1.29539899983609	0.817233353\\
};
\addlegendentry{Port 1: $\eta^+_0$};

\addplot [color=red,solid,line width=1.3pt]
  table[row sep=crcr]{%
1.00003886164028	0.78691993\\
1.01257577144587	0.786702593\\
1.0229987394275	0.803153392\\
1.03669290599983	0.816515259\\
1.04952662306364	0.82432739\\
1.06157877160113	0.830889728\\
1.07635051623125	0.835318208\\
1.0893956017647	0.839254343\\
1.10295576893376	0.845310125\\
1.11604138041654	0.848894819\\
1.12927345508248	0.851911248\\
1.14292447935682	0.854302894\\
1.15617514065785	0.85646531\\
1.17529087784906	0.861493593\\
1.18974703772829	0.863302331\\
1.2052370259554	0.864657927\\
1.221338185652	0.865260484\\
1.23808341607765	0.865408022\\
1.25865764038518	0.865207939\\
1.27656684526565	0.864231678\\
1.29539899983609	0.862319939\\
};
\addlegendentry{Port 2: $\eta^-_0$};

\end{axis}
\end{tikzpicture}%
	\caption{\color{black} Radiation efficiency at the broadside frequency versus transversal asymmetry, $T$, at 24 GHz, obtained by full-wave simulation for 8 unit cells. For high TA, i.e. \mbox{$T\gg 1$}, the radiation efficiencies in the forward and backward directions correspond to the efficiencies of the series \mbox{$\eta_{se}$} and shunt resonators \mbox{$\eta_{se}$}, respectively.\color{black}}
	\label{fig:eta_vs_T}
\end{figure}
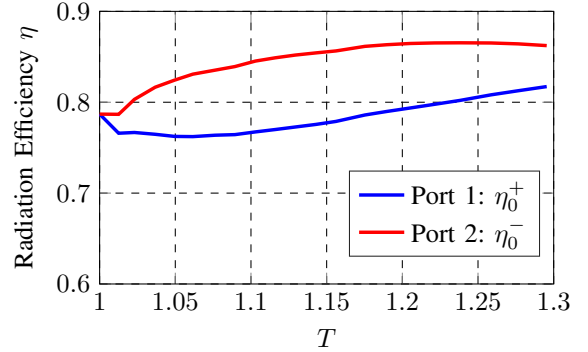

For \mbox{$T=1$}, corresponding to the far-left point in Fig.~\ref{fig:D_vs_T}, the unit cell is longitudinally asymmetric but transversely symmetric. In this case, the polarization is generally elliptic, and circular for the proper amount of LA~\cite{Otto_TAP_04_2014}, as in the design of Fig.~\ref{fig:D_vs_T} where the directivities at broadside are equal for both polarizations \color{black}($v$ and $h$)\color{black}. Since the excitation ports are oppositely located in space along the axis of the antenna, the directions of circular polarization are mutually opposite in the two directions, being LHCP in one direction and RHCP in the other direction.

\color{black}
Figure~\ref{fig:eta_vs_T} plots the broadside radiation efficiencies \mbox{$\eta^+_0$} and \mbox{$\eta^-_0$} when the antenna is excited in the forward and backward directions, respectively. For \mbox{$T=1$}, i.e. in the LA case, the two radiation efficiencies are equal, as is always the case when the unit cell is transversally symmetric. Furthermore, the two radiation efficiencies increase and tend to separate the forward and backward efficiencies for increasing transversal asymmetry. This fact is supported by Fig.~\ref{fig:Jsurf_SFP}, where the surface current densities correspond to the series and shunt modes when the antenna is excited in the forward and backward directions, respectively. For $T\approx 1.3$ the radiation efficiencies are $\eta^+_0=86~\%$ and $\eta^-_0=81~\%$.
\color{black}

\begin{figure}
	\centering
	\hspace{5mm}
	\setlength\figureheight{.3\textwidth}
	\setlength\figurewidth{.35\textwidth}
	\input{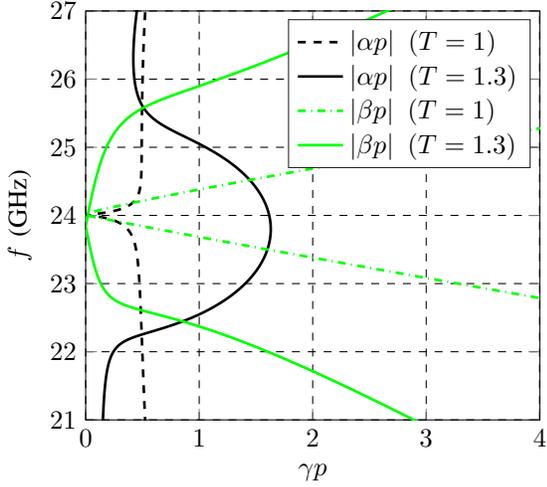}
	\caption{Complex propagation constant $\gamma p$ for DS $(T=1)$ and DA $(T=1.3)$ unit cells with period $p$. The DA exhibits high loss factor $\alpha$ at the transition frequency $f_0=24~\textrm{GHz}$. }
	\label{fig:gamma}
\end{figure}

Introducing TA and progressively increasing $T$, and thus progressively departing from the equal power splitting regime, results in transforming circular polarization into elliptic polarization with progressively increasing ellipticity, until the polarization becomes essentially linear. So, the cross-polarization level is directly controlled by the degree of TA, with higher $T$ corresponding to lower cross-polarization. At $T\approx 1.3$, the cross-pol discrimination reaches about 19~dB. The decrease in the directivity may also be described in terms of an increase in the loss factor ($\alpha$) at broadside, with corresponding decrease in the effective aperture of the P-LWA, as shown in Fig.~\ref{fig:gamma}.

\begin{figure}
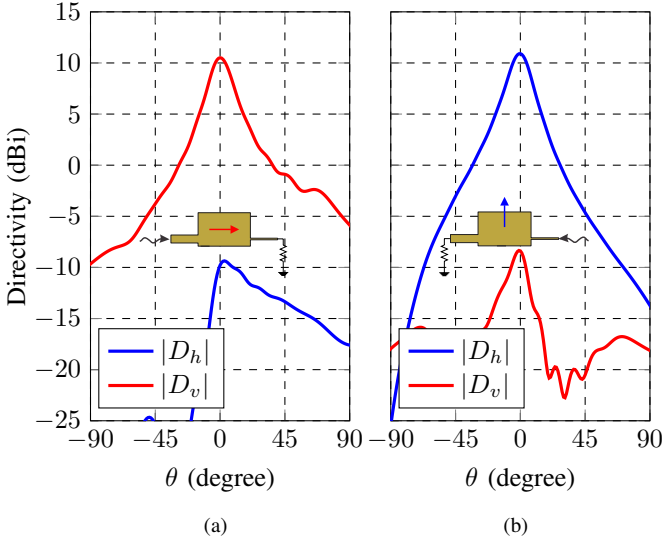

	\centering
	\setlength\figureheight{.3\textwidth}
	\setlength\figurewidth{.19\textwidth}
	\mbox{
	\hspace{6mm}
	\subfigure[]{\input{SFP_SimDir_Port1.tikz}\label{fig:SFP_FF1}}
	\hspace{3mm}
	
	\subfigure[]{\input{SFP_SimDir_Port2.tikz}\label{fig:SFP_FF2}}
	}
	\caption{Full-wave simulated directivity (8 unit cells as in Fig.~\ref{fig:D_vs_T}) \textcolor{black}{for $\phi = 90^\circ$}. (a)~Shows longitudinal (horizontal) and (b)~transversal (vertical) polarizations corresponding to the series and shunt radiation mechanisms, respectively.}
	\label{fig:SFP_Dir_FF}
\end{figure}

Finally, the full-wave simulated forward and backward directivities of the antenna are shown in Fig.~\ref{fig:SFP_Dir_FF}. A good level of cross-polarization discrimination is observed over a wide range of radiation angles.

\subsection{Series-Fed Capacitively-Coupled Patch (SF-CCP) P-LWA}\label{sec:CSFP}

\begin{figure}
	\centering
	\includegraphics[width=.48\textwidth]{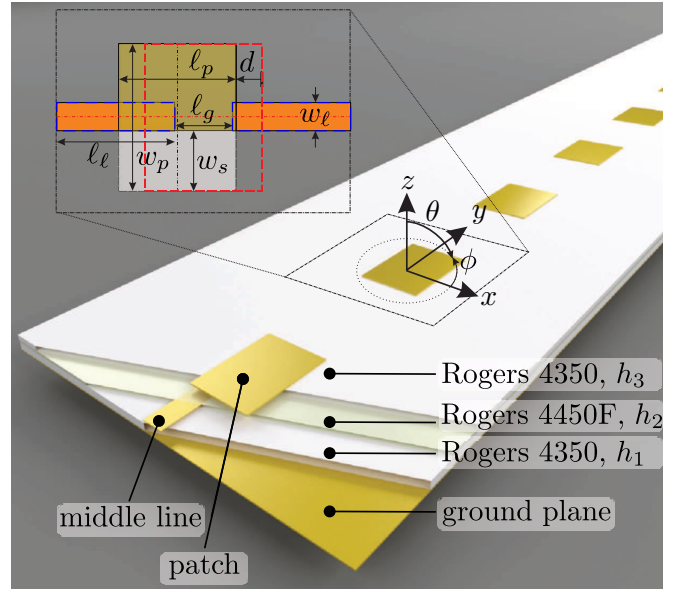}
	\caption{Series-fed coupled patch (SF-CCP) double asymmetric P-LWA. The corresponding double symmetric structure is shown by continued dashed lines. Symmetry w.r.t. the longitudinal axis is broken by vertically cutting out the patch by a quantity $w_s$, while symmetry w.r.t. the transverse axis has been broken by shifting the patch in the longitudinal direction by a quantity $d$.}
	\label{fig:CSFP_layout}
\end{figure}

The SF-CCP P-LWA, which also radiates in the $n = −1$ space harmonic, is shown in Fig.~\ref{fig:CSFP_layout}. The structure includes three metallization layers, supporting respectively an array of patches (top layer), an array of gapped transmission lines (middle layer) and the ground plane (bottom layer). This P-LWA is based on the proximity-coupled patch antenna design reported in~\cite{Chung_06_1993} and later extended to a P-LWA in~\cite{Al-Bassam_GeMIC_03_2012}. The structure includes two Rogers RO~4350 substrate laminates cores with heights $h_1=254~\mu\textrm{m}$ and $h_3=168~\mu\textrm{m}$, respectively. The two substrates are stacked together with a Rogers RO~4450F prepreg laminate of height $h_2=100~\mu\textrm{m}$. The geometrical dimensions are indicated in Fig.~\ref{fig:CSFP_layout}. The same design procedure as in Sec.~\ref{sec:SFP} is used. The geometrical lengths $\ell_\ell$ and $\ell_p$ are tuned to obtain frequency-balancing condition $\omega_\text{se}=\omega_\text{sh}=\omega_0$ and the $Q$-balancing condition is obtained by trimming the patch on the top metallization layer by a quantity $w_s$ w.r.t. the longitudinal axis. Here, TA, and therefore DA, is obtained by shifting the patch in the longitudinal direction by a quantity $d$, which has the same effect on the Bloch impedance and directivity at broadside as the variation of the parameters $w_{\ell1}$ and $w_{\ell2}$ in the SF-LCP P-LWA of Sec.~\ref{sec:SFP}. The geometrical dimensions are given in Tab.~\ref{tab:CSFP_dim}.

\begin{table}
	\centering
		\caption{Dimensions in mm of the DA CSFP P-LWA shown in Fig.~\ref{fig:CSFP_layout}.}
		\begin{tabular}{c|c|c|c|c|c|c}
			$l_{l}$ & $l_{p}$ & $l_g$ &$d$ &$w_{l}$ & $w_{p}$ & $w_{s}$  \\ \hline \hline
			2.93 & 2.95 & 1.44 & 0.65 & 0.70 & 3.60 & 1.45
		\end{tabular}
	\label{tab:CSFP_dim}
\end{table}

\begin{figure}
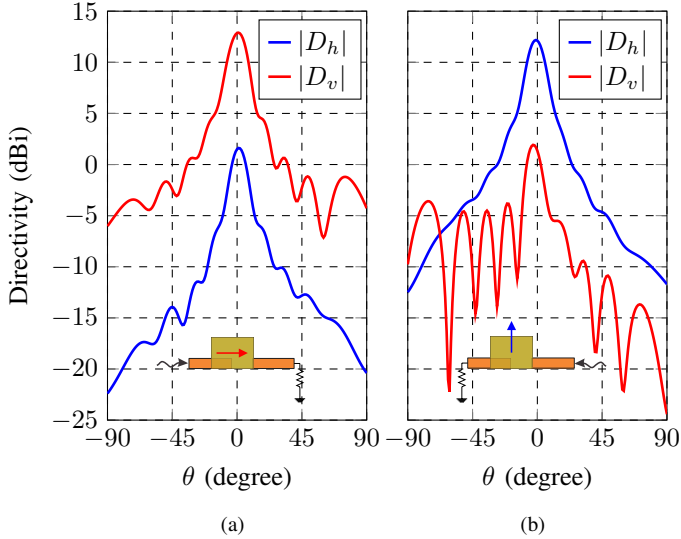

	\centering
	\setlength\figureheight{.3\textwidth}
	\setlength\figurewidth{.19\textwidth}
	\mbox{
	\hspace{6mm}
	\subfigure[]{\input{SFCP_SimDir_Port1.tikz}\label{fig:SFCP_FF1}}
	\hspace{3mm}
	
	\subfigure[]{\input{SFCP_SimDir_Port2.tikz}\label{fig:SFCP_FF2}}
	}
	\caption{Full-wave simulated directivity of the DA P-LWA shown in Fig.~\ref{fig:CSFP_layout} with 8 unit cells \textcolor{black}{for $\phi = 90^\circ$}. (a)~Longitudinal (horizontal) and (b)~transverse (vertical) polarizations corresponding to the series and shunt radiation mechanisms, respectively.}
	\label{fig:SFCP_Dir_FF}
\end{figure}

The full-wave simulated directivities are shown in Fig.~\ref{fig:SFCP_Dir_FF}. The results are qualitatively identical to those obtained in Fig.~\ref{fig:SFP_Dir_FF} for the DA SF-LCP P-LWA.

\color{black}Finally, the design parameters of both the DA SF-LCP and SF-CCP P-LWAs and their circuit counterparts are summarized in Tab.~\ref{tab:para_summary}.\color{black}

\begin{table}
	\centering
		\color{black}
		\caption{Summary of the geometrical and circuit parameters influences in the four symmetry cases.}
		\color{black}
		\begin{tabular}{l|c|c}
			Symmetry case & Geometrical parameter & Circuit l parameter \\ \hline \hline
			DS, LA, TA, DA & $l_p$, $l_l$ & $f_\text{se}$, $f_\text{sh}$ \\
			LA, TA, DA		 & $w_s$				& $Q_\text{se}$, $Q_\text{sh}$ \\
			TA, DA				 & $\Delta w = \left|w_{l1}-w_{l2}\right|$ or $d$ & $T$
		\end{tabular}
		\color{black}		
	\label{tab:para_summary}
\end{table}

\section{Transmission Line Matching Network}\label{sec:matching}

The proposed DA P-LWA is intended to operate broadside. Therefore, the matching networks are designed at the corresponding frequency (24~GHz). Matching is presented here only for the SF-LCP P-LWA, that will be experimentally demonstrated in the next section.

\begin{figure}[htbp]
	\centering
	\subfigure[]{\includegraphics[width=.48\textwidth]{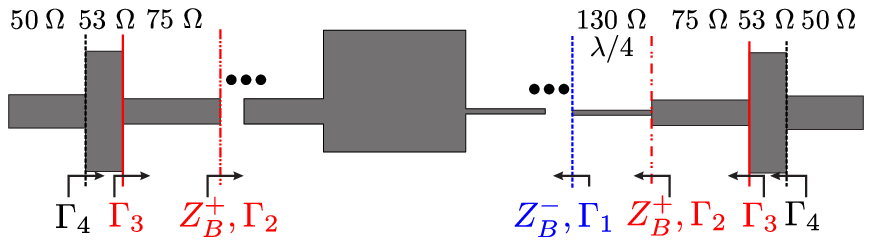}\label{fig:MN_layout}}
	\subfigure[]{\includegraphics[]{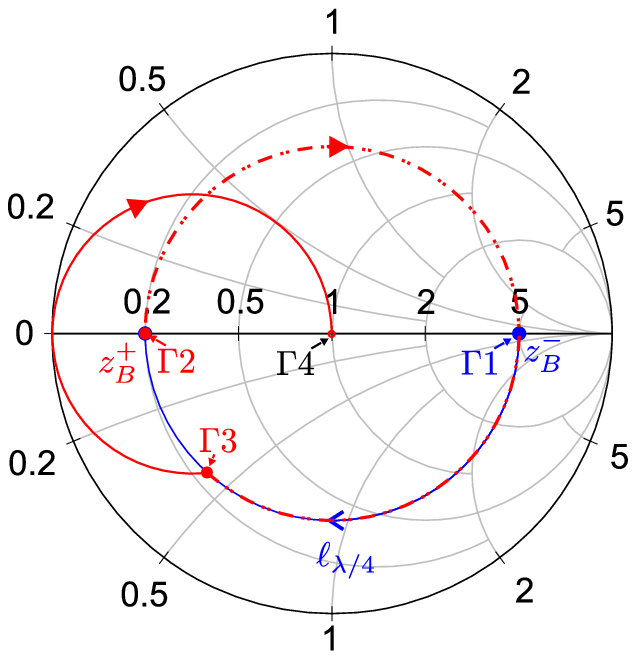}\label{fig:SC}}
	\caption{Matching network for the SF-LCP P-LWA. (a) Geometry and parameters of the forward and backward matching network. (b) Smith Chart.}
	\label{fig:MN}
\end{figure}

It was found that using simple quarter-wavelength transmission-line transformers was not appropriate for matching because the corresponding characteristic impedances were too low at Port~1 and too high at Port~2, respectively. Instead, matching is accomplished using double-sided open-ended stub as shown in Fig.~\ref{fig:MN_layout}.

At both ends, the Bloch impedance is resonant and therefore purely real at broadside (see Fig.~\ref{fig:ZB_SFP}), which places the reference planes of the antenna, $z_\text{B}^+$ and $z_\text{B}^-$\footnote{$z_\text{B}^+$ and $z_\text{B}^-$ are the normalized forward and backward Bloch impedances to the trasversally symmetric Bloch impedance $Z_{\text{B},0,\textrm{sym}}=\sqrt{R/G}$, respectively},  on the real axis of the Smith chart, as shown in Fig.~\ref{fig:SC}. At Port~1, $z_\text{B}^+$ is matched through a transmission-line section followed by the stub. At Port~2, $z_\text{B}^-$ is first brought to $z_\text{B}^+$ using a quarter-wavelength transformer of appropriate characteristic impedance (130~$\Omega$), so that the same line-stub transformer as at Port~1 is then used.

\section{Experimental Demonstration}\label{sec:exper}

\begin{figure}
		\includegraphics[width=0.48\textwidth]{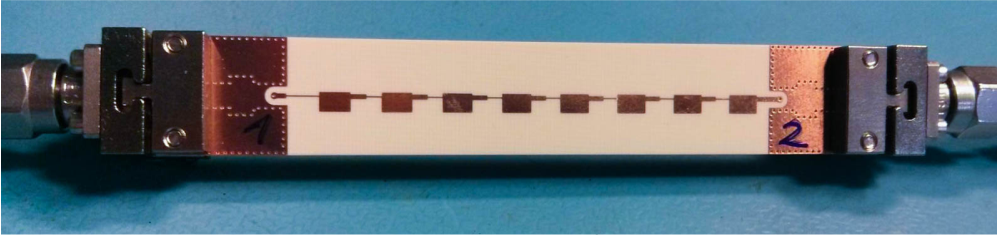}
		\caption{Top view of the fabricated prototype.}
		\label{fig:proto}
\end{figure}

Figure~\ref{fig:proto} shows the fabricated prototype. The feeding networks are placed at the back of the structure with metallized via transitions to the antenna and end-launch connectors to the ports.

\begin{figure}
		\setlength\figureheight{.25\textwidth}
		\setlength\figurewidth{.38\textwidth}
		\centering
		\input{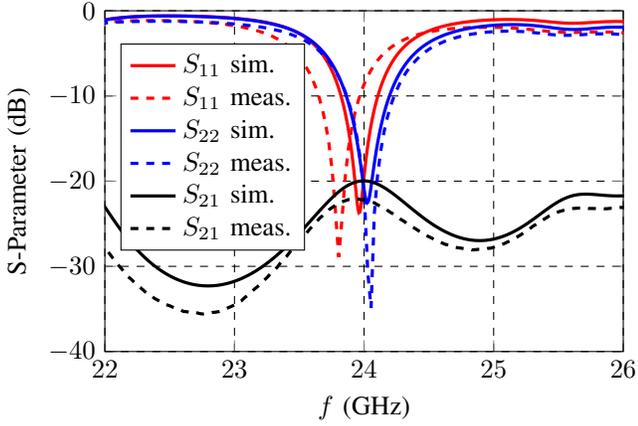}
		\caption{Measured and simulated S-parameters. $S_{12}=S_{21}$ by reciprocity.}
		\label{fig:spar}
\end{figure}

Figure~\ref{fig:spar} compares the measured and simulated S-parameters, verifying that good matching is achieved at the design frequency around 24~GHz. The slight deviations in the experiment are attributed to fabrication errors. \color{black} Furthermore, the low level of the transmission coefficient, $S_{21}$, clearly indicates high radiation level, as expected from Fig.~\ref{fig:gamma}. Additionally, the relative frequency bandwidth of the P-LWA may be evaluated from the reflection coefficients, $S_{11}$ and $S_{22}$, which are both approximately to $1.3~\%$. The reason for the rather small bandwidth is the specific design of the matching/feeding section.
\color{black}

\begin{figure}
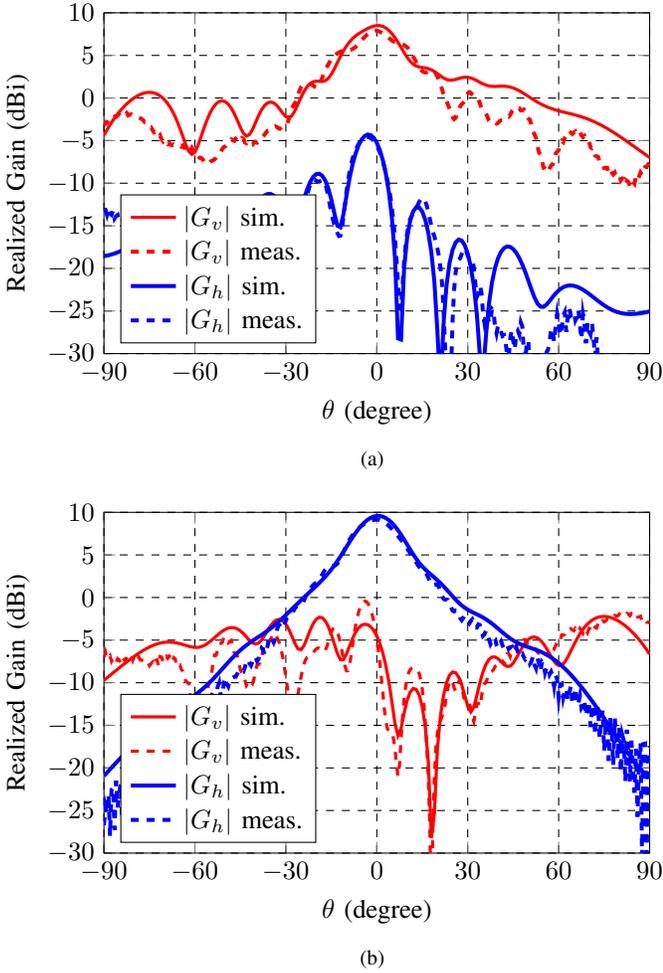

		\setlength\figureheight{.25\textwidth}
		\setlength\figurewidth{.4\textwidth}
		\centering
		\hspace{12mm}
		\subfigure[]{\input{SimMeas_Gain_Port1.tikz}\label{fig:gain1}}
		
		\hspace{12mm}
		\subfigure[]{\input{SimMeas_Gain_Port2.tikz}\label{fig:gain2}}
		\caption{Measured and simulated radiation patterns in the principal radiation plane ($yz$-plane). (a)~Excitation at Port~1 and (b)~excitation at Port~2.}
		\label{fig:gain_compare}
\end{figure}

Figure~\ref{fig:gain_compare} shows the measured and FDTD-simulated realized gain of the antenna for excitation at Port~1 [Fig.~\ref{fig:gain1}] and at Port~2 [Fig.~\ref{fig:gain2}], while port~2 and port~1 are terminated with a match load, respectively. \color{black}The realized gain of port~1 and port~2 at broadside are 8.5~dBi and 9.5~dBi and the cross-polarization discrimination levels are $13.75~\text{dB}$ and $14~\text{dB}$ , respectively\color{black}. Excellent agreement is observed between the simulation and measurement results, both confirming the dual-channel orthogonal-polarization operation of the P-LWA.

\color{black}
Note that the gain of this specific DA P-LWA cannot be increased by increasing the length of the antenna, i.e. cascading more unit cells, because of the low power left at the end of the structure. To overcome this issue, a different unit-cell design could be employed with lower radiation/loss factor $\alpha$ at broadside.\color{black}

\section{Conclusion}\label{sec:conc}

Unit-cell asymmetries play a fundamental role in the operation of P-LWAs. We have shown here that double-asymmetry, i.e. asymmetry with respect to both the longitudinal and transverse directions of the antenna structure, creates an interesting regime where the P-LWA can simultaneously radiate two orthogonal waves when excited from its two ports. This is a new feature in P-LWAs that both completes the symmetry investigation of such antennas and provides a novel technology for polarization diversity. The dual-channel orthogonal-polarization operation of the DA P-LWA has been demonstrated by full-wave and experimental results.

\color{black}
\appendices

\section{Bloch Impedance Derivation and Analaysis}\label{appen:network}

This section recalls the derivation of the formulas for the Bloch impedance in terms of the the transformation ratio, $T$~\cite{Otto_TAP_10_2014}, since these formulas constitute the foundation of the paper.

The lattice circuit model in Fig.~\ref{fig:lattice}, consisting of the two series and two shunt resonators that can be isolated by odd and even excitations~\cite{Otto_TAP_10_2011}, respectively, cannot describe TA and hence DA, due to the absence of coupling between the two types of resonators. Therefore, one has to resort to Fig.~\ref{fig:lattice_t}, where this coupling is modeled by four ideal transformers with transformation ratio $T$~\cite{Otto_TAP_10_2014}.

\begin{figure}
		\includegraphics[width=0.48\textwidth]{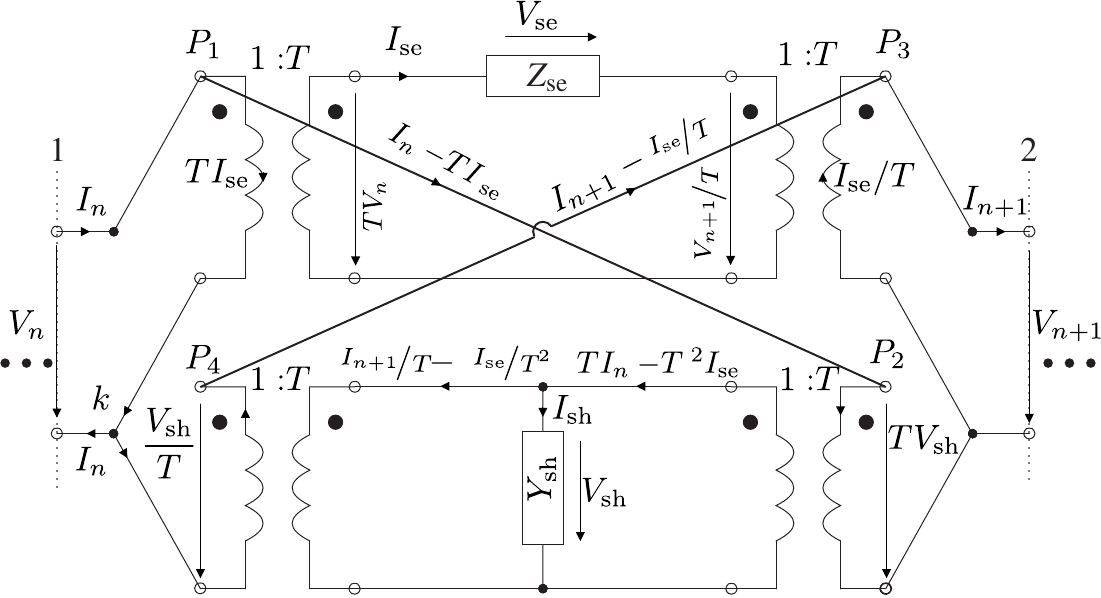}
		\caption{\color{black} General P-LWA circuit model of Fig.~\ref{fig:lattice_t} with voltage, current and reference point notations required for the derivation of the Bloch impedance.\color{black}}
		\label{fig:lattice_t_appen}
\end{figure}

In the forthcoming derivations, we shall consider the voltage, current and reference point notations in Fig.~\ref{fig:lattice_t_appen}. Given periodicity, the voltage and current at the input and output terminals of the circuit are related by the Floquet-Bloch theorem, i.e.
\begin{align}
	V_{n+1} = V_{n}e^{-\gamma p}\label{eq:Vn},\\
	I_{n+1} = I_{n}e^{-\gamma p}\label{eq:In},
\end{align}
where $\gamma$ and $p$ are the propagation constant and period, respectively, of the structure. The Bloch impedance is defined as $Z_\text{B}=V_n/I_n$, $\forall n$, and we wish to express it here in terms of $T$, $Z_\text{se}$ and $Y_\text{sh}$.

We first apply the Kirchhoff's voltage law to the loop $P_1-P_2-P_3-P_4-P_1$. Doing so and using~\eqref{eq:Vn} yields

\begin{align}\label{eq:Vsh_Vn}
	V_\text{sh}\left(T+\dfrac{1}{T}\right) = V_n\left(1+e^{-\gamma p}\right),
\end{align}

\noindent where $V_\text{sh}$ is the voltage drop across the shunt resonator. Applying next the Kirchhoff's current law to node $k$ and using~\eqref{eq:In} similarly yields

\begin{align}\label{eq:Ise_In}
	I_\text{se}\left(T+\dfrac{1}{T}\right) = I_n\left(1+e^{-\gamma p}\right).
\end{align}

The Bloch impedance is therefore simply obtained by dividing~\eqref{eq:Vsh_Vn} by~\eqref{eq:Ise_In}, which yields

\begin{align}\label{eq:ZB_Vsh_Ise}
	Z_\text{B}=\dfrac{V_n}{I_n}=\dfrac{V_{n+1}}{I_{n+1}}
=\dfrac{V_\text{sh}}{I_\text{se}}, 
\end{align}

\noindent where we next need to express $V_\text{sh}/I_\text{se}$ in terms of $T$, $Z_\text{se}$ and $Y_\text{sh}$. This may be achieved by applying Kirchhoff's voltage law to the $Z_\text{se}$ loop and Kirchhoff's current law to the $Y_\text{se}$ node, resulting into

\begin{align}\label{eq:Ise_Vn}
	I_\text{se}Z_\text{se} = TV_n - \dfrac{V_{n+1}}{T},
\end{align}

and

\begin{subequations}
\begin{align}\label{eq:Vsh_Ise_In}
	V_\text{sh}Y_\text{sh} = TI_n-\dfrac{I_{n+1}}{T}-2I_\text{se}\tau,
\end{align}
where
\begin{equation}
\tau=\dfrac{1}{2}\left(T^2-\dfrac{1}{T^2}\right).
\end{equation}
\end{subequations}

\noindent In~\eqref{eq:Vsh_Ise_In}, $I_n$ and $I_{n+1}$ may be expressed in terms of $Z_\text{B}$, $V_n$ and $I_{n+1}$ using~\eqref{eq:ZB_Vsh_Ise}, which yields

\begin{align}\label{eq:ZB_Vsh_tau}
		V_\text{sh}Y_\text{sh} = \dfrac{1}{Z_\text{B}}\left(TV_n-\dfrac{V_{n+1}}{T}\right)-2I_\text{se}\tau.
\end{align}

\noindent Noticing that the bracket term in this expression is exactly the right-hand side of~\eqref{eq:Ise_Vn}, this relation transforms to

\begin{align}\label{eq:Vsh_Ise_tau}
		V_\text{sh}Y_\text{sh} = \dfrac{I_\text{se}Z_\text{se}}{Z_\text{B}}-2I_\text{se}\tau,
\end{align}

\noindent Upon division by $I_\text{se}$, Eq.~\eqref{eq:Vsh_Ise_tau} provides the required $V_\text{sh}/I_\text{se}$ ratio in~\eqref{eq:ZB_Vsh_Ise}. We obtain thus, after rearranging,

\begin{align}\label{eq:ZB_quad}
	Z_\text{B}^2 + \dfrac{2\tau}{Y_\text{sh}}Z_\text{B}-\dfrac{Z_\text{se}}{Y_\text{sh}} = 0,
\end{align}

\noindent which is a quadratic equation in $Z_\text{B}$, with solution

\begin{align}\label{eq:ZBpm}
Z_\text{B}^\pm= \mp \dfrac{\tau}{Y_\text{sh}} + \sqrt{\left(\dfrac{\tau}{Y_\text{sh}}\right)^2+\dfrac{Z_\text{se}}{Y_\text{sh}}}.
\end{align}

We shall next show that the transformation ratio, $T$ or $\tau$, controls the Bloch impedance at broadside only. For this purpose, we insert~\eqref{eq:zse_ysh} into~\eqref{eq:ZBpm}, and evaluate the asymptotic result in the off-broadside regime, which may be considered to correspond to $\Delta\omega=\omega-\omega_0\rightarrow\infty$~\cite{Otto_TAP_10_2014}:

\begin{align}\label{eq:ZB_offbroad}
	 \lim_{\Delta\omega \to \infty} Z_\text{B}^\pm  = & \lim_{\Delta\omega \to \infty}  \mp \dfrac{\tau}{G + j2C\Delta\omega} \nonumber \\
	& + \sqrt{\left(\dfrac{\tau}{G + j2C\Delta\omega}\right)^2+\left(\dfrac{R + j2L\Delta\omega}{G + j2C\Delta\omega}\right)} \nonumber \\
	 = & \sqrt{\dfrac{L}{C}} = Z_{\text{B},\Delta},
\end{align}

\noindent where the subscript $\Delta$ refers to the off-broadside operating regime. Note that $Z_{\text{B},\Delta}$ does \emph{not} dependent on $\tau$, and hence on $T$, so that, its insertion into~\eqref{eq:eta_pm} yields the same result for $\eta^+_\Delta$ and $\eta^-_\Delta$. This proves that $T$ affects the P-LWA properties only at broadside, according to~\eqref{eq:ZBpm}.

Finally, the Bloch impedance at broadside may be derived by imposing $\Delta\omega=0$~\eqref{eq:ZBpm}, which results into\eqref{eq:ZB}.

Tuning $T$ in a practical P-LAW may be easily achieved by changing the geometry features around the transverse axis, as indicated in Fig.~\ref{fig:DA_def}. In reality, the response of the P-LWA is very sensitive to geometrical parameter changes; even a deeply sub-wavelength variation induces a substantial difference in the Bloch impedance and hence in the antenna efficiency.
\color{black}

\ifCLASSOPTIONcaptionsoff
  \newpage
\fi
\bibliographystyle{IEEEtran}
\bibliography{IEEEabrv,TAP_2016_Double_Asymmetry_in_LWAs_al-bassam}

\end{document}